\documentstyle [titlepage,12pt,seceqn]{article}
\begin{document}
\pagestyle{plain}
\baselineskip=0.62cm
\def\be {\begin{equation}}
\def\ee {\end{equation}}
\def\bd {\begin{displaymath}}
\def\ed {\end{displaymath}}
\begin{flushright}
{Crete$-97-19$}\\
\end{flushright}
\vspace{1.5cm}
\begin{center}
{\large
Vortex Pull by an External Current 
}
\end{center}
\vspace {1.2cm}

\centerline{ {\large G.N. Stratopoulos  \footnote{
email: stratos@physics.uch.gr} } }
\vskip 0.5cm

\centerline {\it {Department of Physics and Institute of Plasma Physics,
 University of Crete,}}
\centerline  {\it {and Research Center of Crete,}}
\centerline {\it {P.O.Box 2208, 710 03 Heraklion, Crete; Greece }}
\vskip 1.5cm
\centerline{\large\bf Abstract}

In the context of a dynamical Ginzburg-Landau model it is shown 
numerically that under the influence of a homogeneous external 
current {\bf J} the vortex drifts against the current with velocity 
${\bf V}= -{\bf J}$ in agreement to earlier analytical predictions. 
In the presence of dissipation the vortex undergoes skew deflection 
at an angle $90^{\circ} < \delta < 180^{\circ}$ with respect to the 
external current. It is shown analytically and verified numerically 
that the angle $\delta$ and the speed of the vortex are linked through 
a simple mathematical relation.

\newpage

\section{Introduction}

In a recent publication \cite{rPTb} it was proposed to study the dynamics
of isolated Abrikosov vortices in the framework of a phenomenological 
effective field theoretic model \cite{rFeynman},
which alternatively can be viewed as a time dependent version of the 
Ginzburg-Landau equation (TDGL). 
Therein, it was possible to derive analytically the equation of motion 
of the guiding center of the vortex (Hall equation), under the influence 
of any kind of external force \cite{rST}.
According to the latter, the guiding center of the vortex moves in a 
direction perpendicular to the externally applied force in analogy to the
planar motion of charged particles in the presence of a perpendicular
magnetic field. The importance of the Hall equation stems from theoretical
considerations which suggest that the guiding center describes quite 
accurately the ``mean'' position of the vortex, a suggestion which was 
verified and quantified by a recent numerical study \cite{rST2}.
It is thus possible to obtain useful information concerning the motion of 
a vortex by simply invoking Hall equation. For instance, by  plain 
implementation of the latter one finds that in the presence of a 
homogeneous external current the vortex drifts against the current.

By construction, the Hall equation is limited to the description of the 
gross features of the motion of the vortex. On the other hand,
an understanding of the finer details of its motion requires 
a detailed solution of the TDGL equations. Such an undertaking has to
rely on numerical methods, due to the nonlinear nature of the
relevant equations. The main purpose of this paper is to report on a 
numerical study of the dynamics of vortices within this model.
In particular we will study the response of a vortex to an external 
current with or without dissipation, an issue of obvious interest in the 
context of superconductivity \cite{rHueb},\cite{rVinokur}.

The paper is organized as follows.
Section 2 contains a general introduction to the model. Its relevance 
to the physics of the superconductor is commented upon, and the main
theoretical predictions concerning the motion of the vortex are
illustrated. Finally, it is shown that an alternative interpretation
of Magnus force \cite{rMagnus} arises 
naturally in terms of the Hall equation.
In Section 3 we incorporate the effect of dissipation and that of an
external electric current in our field theoretical formalism
and we derive the equation of motion for the vortex through
explicit calculations. 
The results of an extensive  numerical study are presented in Section 4.
Vortices are found to behave in accordance to our earlier theoretical 
predictions. As a byproduct of this study we investigate the detail
of the internal ``Cyclotron motion'' of vortices and we show
that an isolated vortex is spontaneously pinned.
In the concluding Section 5 we discuss the relevance of our results 
to the actual superconductor and we propose more realistic 
3-dimensional studies.

\section{The Model}

   Most of our techniques and conventions are described
elsewhere \cite{rPTb},\cite{rST2}, so we briefly  outline 
here the physical and mathematical background of the theory.
   The model in question was originally introduced by  Feynman 
\cite{rFeynman} as a natural dynamical extension of the
Ginzburg-Landau static theory of superconductivity. 
   Indeed, by attributing the correct physical content to fields and 
parameters, it becomes a rather realistic phenomenological model
of a superconductor \cite{rST2}.
   The model admits infinitely long, smooth, cylindrically symmetric flux 
vortex solutions, whose static properties together with the properties 
of pairs of them, have been studied in detail in reference \cite{rST}.
   Our objective is to study some aspects of vortex dynamics within
this model. In doing so, we will ignore excitations along the axis
of the vortex (which by convention is taken to be parallel to z-axis)
and we will directly define the model in two space dimensions.

  As usual, the important dynamical variable is a complex order
parameter $\Psi$ which may be thought as an electrically charged
field coupled to the electromagnetic potential $(A_0,{\bf A})$.
The fields  satisfy the coupled system of TDGL equations
(to simplify notation, fields and coordinates are rescaled along 
the lines of reference \cite{rST2})
\vspace*{0.5cm}
\begin{displaymath}
i \dot \Psi=-{1\over 2} {{\bf D}\sp
2} \Psi + A_0 \Psi+{1\over 4}
{\kappa}^2 (\Psi\sp\ast  \Psi-1)\Psi
\end{displaymath}
\begin{equation}
{1\over \beta}\dot E_i=\epsilon_{ij} \partial_j B - J^s_i
\;\;\;\;\;\;\;\;\;\;\;
{1\over \beta} \;\partial_i E_i\;=\; \rho 
\label{rescaledeqs}
\end{equation}

\vspace*{0.5cm}
\noindent
with $\, B = \epsilon_{ij} \partial_i A_j$,
$ \, E_i = -\partial_t A_i - \partial_i A_0 $,
$ \, D_i=\partial_i-i A_i$, $\beta$, $\kappa$ are free parameters, and 
the spatial indices $i,j$ range from 1 to 2.
   The supercurrent density $\, {\bf J}^s \,$ and the charge density $\, 
\rho \,$ are given by $ J^s_i={1\over {2i}} [\Psi \sp\ast D_i\Psi-c.c]$
and $ \rho=\Psi\sp\ast \Psi - 1$. 
Note that in order to allow the possibility of a condensate ($|\Psi| 
= 1$) at infinity we introduce a background (positive-ion) charge density 
$\, \rho_b \,$ to neutralize the system. For simplicity $\, \rho_b \,$ is 
taken to be constant and homogeneous throughout i.e. 
$\, \rho_b=-\rho_s$ where $\rho_s$ is defined as 
$\rho_s \equiv \Psi^* \Psi|_{\infty}=1$.

The nonlinear system of equations (\ref{rescaledeqs}) admits
static, axially symmetric localized vortex solutions. 
They are similar in nature to the well known Abrikosov vortices 
of the static GL theory. Moreover, in a certain region of the 
parameter space their characteristic lengths (penetration depth,
coherence length) fall well within the scale of a typical type II
superconductor \cite{rST2}.
However, the vortices in this model differ from the ordinary
Abrikosov vortices mainly in one respect; although they carry zero 
electric charge as a whole, there is a local charge modulation in 
their interior i.e. they have non-vanishing charge density and as 
a result an electric field is associated with them.

A remarkable  feature of the model in hand, is the unusual response
of the vortices to external probes.
Indeed, as it was shown in references \cite{rPTb},\cite{rST},
the motion of the vortex is governed by the equation
\begin{equation}
V_{L\,i}=-{1\over {2 \pi N}} \;\epsilon_{ij} \;F_j
\label{Halleqs}
\end{equation}
\noindent
where ${\bf V}_L$ is the mean velocity of the vortex and ${\bf F}$ is 
the sum of the external forces acting upon it. $N$ is the well-known 
integer-valued winding number or  topological charge characterizing
any finite energy configuration which counts the number of times
the phase of $\, \Psi \,$ rotates around the internal circle as we 
scan the circle at spatial infinity \cite{rSolitons}.
$N$ is a conserved quantity and can be written as
the integral of a properly chosen ``topological density'' $\tau$.
Among other possibilities $\, \tau \, $ may be defined as $\, \tau \,=\,
{1\over 2 \pi } B$, a definition which entails the familiar 
magnetic flux quantization and it is of obvious physical interest
while, for our purposes the most useful form of $\tau$ is
\begin{equation}
\tau\;=\;{1\over 2 \pi i}
\;[\epsilon_{kl} (D_k\Psi)\sp\ast (D_l\Psi)-i B (\Psi\sp\ast \Psi-1)]
\label{TopChargeII}
\end{equation}
\noindent
The importance of the above formula stems from the fact that it appears
in the expressions for the momentum and the angular momentum of the theory.
Indeed, it was pointed out in reference \cite{rPTb}, that the
Noether expression for the linear momentum of the model is ambiguous
for any configuration with non-zero topological charge. The unambiguous
expression turns out to be
\begin{equation}
 P_k=\epsilon_{ki} \int d^2x\; (2 \pi\; x_i\; \tau +
{1\over \beta}\;E_i B)
\label{momentum}
\end{equation}
\noindent
which in turn leads to a radical revision
of the physical interpretation of the momentum.
The presence of the first moment of the topological density
in the expression for the  linear momentum, directly
associates the latter with the position of the vortex.
In fact, it is possible to show \cite{rPTb}
that the ``mean position'' of the vortex is described by 
a quantity ${\bf R}$ called the guiding center of the 
configuration and defined as
$\, R_i\equiv -{1\over 2\pi N} \epsilon_{ij} P_j \,$.
By identifying ${\bf R}$ with the position and $\dot {\bf R}$
with the mean velocity ${\bf V}_L$ of the vortex, and assuming 
that a generic force $F_j = {dP_j\over dt} \,$ acts on the system,
we end up with equation (\ref{Halleqs}).

Contrary to intuition and Newtonian Mechanics, equation (\ref{Halleqs}) 
implies that a vortex moves at a constant 
calculable speed in a direction perpendicular to the applied force. 
This kind of behavior is analogous to the planar motion of charge 
particles under the action of a perpendicular magnetic field \cite{rPTa} 
and thus we call it Hall behavior and from now on we will refer to 
eq.(\ref{Halleqs}) as Hall equation.
Vortex dynamical behavior becomes less exotic and Newtonian Mechanics
reestablishes by adopting an alternate interpretation of the Hall equation.
To do so, we rewrite eq.(\ref{Halleqs}) in the form :  
$\, {\bf V}_L \,=\, -{1\over {2 \pi N}} \, {\bf F} \times \hat e_z \, $
where $\, {\bf V}_L \,$ is the velocity of the vortex in the  $x-y$ plane
and  $\, {\bf F} \, $ is the total force acting upon it. Multiplying 
both sides of the former equation by $\, 2 \pi N \, $ and taking the 
cross product by $ \, \hat e_z \, $ we get
\begin{equation}
2 \pi N \, {\bf V}_L \times \hat e_z  \,=\, {\bf F}
\label{Magnuseqs}
\end{equation}
\noindent
According to eq.(\ref{Magnuseqs}), for a vortex which moves 
with constant velocity Newton's Law $\, \sum {\bf F} \,=\,0 \, $
is restored  if we assume that in addition to any other force acting on 
the vortex another `new'  transverse force $\, {\bf F}_T = - 2 \pi 
N {\bf V}_L \times \hat e_z \, $ also acts on it.
Bearing in mind that the total magnetic flux $\phi_0$ equals 
$\, 2 \pi N \,$ we see that $\, {\bf F}_T \,$ has similar form
to the familiar so called `Magnus force' \cite{rMagnus} which is
usually invoked to describe the motion of vortices in the superconductor.
The similarity becomes more obvious in full units where this additional 
force reads ${\bf F}_T=-\rho_s \phi_0 {\bf V}_L \times \hat e_z$.

As we have explained in some previous work \cite{rST},
Hall behavior (and consequently the transverse force ${\bf F}_T\,$) is 
a generic characteristic of soliton dynamics in systems
with non trivial topology and spontaneously broken Galilean invariance,
due for instance to the presence of a crystal lattice, and has 
a clear mathematical origin.
 Yet one would like to have a more physical explanation for the 
appearance of ${\bf F}_T\,$ in eq. (\ref{Magnuseqs}). 
In fact, one can attain such an explanation by attributing the
origin of ${\bf F}_T\,$ to the interaction between the magnetic
flux of the vortex and the internal electric currents which are
generated by the motion of the vortex \cite{rKF}. Specifically, 
let us assume that the vortex is moving with constant velocity 
$\, {\bf V}_L \,$. In our field theory prescription a moving 
vortex with velocity $\, {\bf V}_L \,$ is a field configuration 
$\, \Psi({\bf r-V}_L t) e^{i({\bf V}_L {\bf r}-{1 \over 2} 
|{\bf V}_L|^2 t)}, {\bf A}({\bf r-V}_L t)$ (see ref. \cite{rManton}
for more details on Galilean  boosts in 2-d vortices). Let us now switch  
to the reference frame where the vortex is still. In that frame, the 
background ions of the crystal lattice form a homogeneous current of 
negative charge carriers with charge density $\, -\rho_s \,$ and 
velocity $\, -{\bf V}_L$. This current interacts with the magnetic field 
of the vortex and as a result feels a Lorentz force acting upon it.
Consequently the vortex feels a backreaction force opposite
to the Lorentz force which can be easily computed and is 
found to be exactly the transverse force ${\bf F}_T\, $ mentioned 
above.

\section{External Current and Dissipation}

    To study the response of the vortex to an externally prescribed
current ${\bf J}^{ext}({\bf x},t)$ we simply substitute
${\bf J}^s \to {\bf J}^s + {\bf J}^{ext}$ in eq. (\ref{rescaledeqs}).
Because of the external current the linear momentum (\ref{momentum})
of the system is no longer conserved. A straightforward application 
of the equations of  motion yields
\begin{equation}
{d\over dt} P_k= F^{Lorentz}_k=-\int d^2x \; \epsilon_{kl}
 J^{ext}_l ({\vec x}, t) B({\vec x}, t) \label{lorentzforce}
\end{equation}
Assuming that the external current is uniform throughout the plane
i.e. $\;{\vec J}^{ext}({\vec x},t) =\, {\vec J_0}$, the only space dependent 
quantity on the right hand side of this equation  is the
magnetic field, and its integral, the total magnetic flux,
is equal to $\, 2\pi \,N $.
Thus, the equation of motion for the momentum takes the simpler form
$\, \dot P_k \,=\, - 2\pi N \epsilon_{kl} J_{0l}$.
Correspondingly the time evolution of the guiding center or the
``mean position'' of the vortex  reads

\begin{equation}
 {dR_k\over dt}\,= \, -{1\over 2\pi N} \epsilon_{ij} \dot P_j
\,= \, - J_{0k}  \label{guidcen2}
\end{equation}

\noindent
which quite surprisingly implies that the vortex drifts against 
the current with constant calculable speed.
Note that in a similar model, Manton \cite{rManton} arrives to the same  
result by considering Galilean boosts on vortex configurations.

The pull of the vortex by the current has a simple explanation in the 
context of Hall equation. The magnetic flux of the vortex exerts on 
the electric current a Lorentz force in the $-90^{\circ}$ direction with 
respect to the current. Consequently the vortex feels a backreaction
force in the $+90^{\circ}$ direction and naively one would expect the 
vortex to move in a direction perpendicular to the current. However,
as it follows from the Hall equation the vortex moves at $+90^{\circ}$ 
with respect to the applied force, and therefore, the 
vortex moves in a direction opposite to the external current.

The introduction of dissipation in the system is a complicated task.
Formally the effect of dissipation in the system is studied by 
adding a, phenomenological friction term in the TDGL equations. 
However, no such term has been derived on the basis of solid 
physical reasoning. Yet, one must have in mind that there are several
restrictions in the form and the properties of any such term.
Any friction term inserted in (\ref{rescaledeqs}) should meet the 
following conditions : a) it should vanish for any static vortex 
solution, b) it should decrease the total energy 
$W$ of the system i.e. $ \,- dW/dt \,$ should be positive definite
and finally c) should preserve the electromagnetic
$U(1)$ gauge invariance of the system,
or equivalently it should preserve the continuity equation.
Our choice, though not unique was the most natural among a small set 
of candidates, and that was to add a term of the form 
$\, C_d \, \epsilon_{ij} \partial_j \dot B$ -with $\, C_d \, $ 
a positive constant- on the right hand side of the equation of motion 
for the electric field in (\ref{rescaledeqs}).

By construction the friction term is gauge invariant and vanishes for 
any static configuration. The time derivative of the energy 
reads $\, dW/dt \, = \,{1\over \beta} \int d^2x \{ E_i \, ( C_d \, 
\epsilon_{ij} \partial_j \dot B) \} \, $ which by integration by parts 
becomes $ \, dW/dt = -{C_d\over \beta} \int d^2x 
\{(\epsilon_{ij} \partial_j E_i) \, \dot B \, \} =
-{C_d\over \beta} \int d^2x \, \dot B^2$. Thus we conclude that the 
friction term we propose meets all the restrictions mentioned above and 
therefore it seems to be a reasonable candidate for a phenomenological 
study of dissipation. 

The time evolution of the linear momentum is
\begin{equation}
{d\over dt} P_k \, = \, C_d \int d^2x \; \{ \epsilon_{kl}
 \, \epsilon_{lm} \partial_m \dot B
  ({\vec x}, t) B({\vec x}, t) \} 
 = \, C_d \int d^2x \, \{ \dot B({\vec x}, t) 
   \partial_k B({\vec x}, t) \}
\label{dissip1}
\end{equation}
\noindent
To proceed further we adopt the rather plausible assumption that in the
presence of a suitably chosen external current as well as dissipation 
a steady state is eventually reached in which the vortex moves rigidly 
with constant velocity ${\vec V}_L$. We assume further that the profile 
of the magnetic field associated with the steady state of the vortex is
of the form $B({\vec x}, t) \, = \, B({\vec x}- {\vec V}_L t ; {\vec V}_L)$.
Then $\dot B({\vec x}, t) \, = \,- V_{Lj} \partial_j B $
and eq. (\ref{dissip1}) reduces to 
\begin{equation}
{d\over dt} P_k \, = \,- S_{kj} V_{Lj}
\hskip 1.3cm with \hskip 1.3cm S_{kj} \,=\, C_d \int d^2x \,
\{ \partial_k B \partial_j B \, \}
\label{dissip2}
\end{equation}
\noindent
A simpler version of this relation is obtained by invoking some further
assumptions about the steady state profile of the vortex, which may be 
viewed as reasonable approximations in the limit of a weak external 
current. Thus, we assume that the vortex retains  approximately its 
initial shape and correspondingly its
axial symmetry i.e. in the rest frame of the vortex the magnetic field
is of the form $\,B=B(\rho) \,$, where $\rho$ is the radial coordinate
in the usual polar variables. Then the diagonal terms of the $S_{kj}$ 
tensor vanish while $S_{11} \,=\, S_{22} = 
C_d/2 \int d^2x \,  (\partial_\rho B^2) \,  \equiv \, \eta  \; $  and 
equation (\ref{dissip2}) reduces to 
\begin{equation}
{d\over dt} P_k \, = \,- \eta V_{Lk}
\label{dissip3}
\end{equation}
\noindent
with $\, \eta \,$ being a positive constant number.
Eq. (\ref{dissip3}) implies that the effect of such a term on an
axially symmetric vortex configuration moving with velocity ${\bf V}_L$
is a force linear to the velocity  of the vortex which
opposes the motion of the latter. We thus believe that with the
insertion of this term we correctly incorporate
the effect of friction in our model.
Finally, the equation of motion of  the guiding center of the vortex
in the presence of both a homogeneous external current and 
dissipation reads
\begin{equation}
V_{Lk} \, \equiv \, {dR_k \over dt} \, = \, - J_{0k}  + {1\over 2\pi N} 
\epsilon_{km} \eta V_{Lm}
\label{guidcen3}
\end{equation}
\noindent
Equation (\ref{guidcen3}) implies that a vortex
in the presence of a transport current, simply drifts against it
 \, (${\bf V_L} = - {\bf J_0}$) ; this leads to a perfect Hall 
effect but with opposite sign to that of the normal-state 
\cite{rDorsey}.  In the presence of dissipation the velocity of the
vortex acquires a component in the perpendicular to the applied current
direction. However, its longitudinal part still has opposite direction
to the transport current, which in turn results in a sign change
of the Hall effect \cite{rDorsey}.  By plain implementation of eq. 
(\ref{guidcen3}) we find that the deflection angle $\delta$ between 
$ {\bf V_L}$ and $ {\bf J_0}$ decreases from $180^{\circ}$ to 
$90^{\circ}$ as 
the contribution of the drag force increases.  Furthermore, a simple 
relation for the deflection angle $\delta$ is obtained, namely
\begin{equation}
cos \delta \,=\, -{\mid {\bf V}_L \mid \over \mid {\bf J}_0 \mid}
\label{goldenrule}
\end{equation}
The former relation becomes obvious by simple inspection of 
fig.1.  There are two forces acting upon the vortex ; 
i) the Lorentz force  
 $\, ( F_k^{Lorentz} \,=\,  2\pi N \epsilon_{kl} J_{0l}) \,$
and ii) the dissipation force $\, ({\bf F}_d \,=\, - \eta {\bf V}_L )$.
According to eq. (\ref{Halleqs}) the vector sum of $\, {\bf F}_d$  and
${\bf F}_{Lorentz}  \,$ rotated by $\, +90^{\circ} \,$ and divided by 
$\, 2 \pi N \, $ must be equal to the velocity of the vortex
$\, {\bf V}_L $. The latter condition leads to relation 
(\ref{goldenrule}). 

From our analysis we conclude that within our model, 
and with or without dissipation the longitudinal part of the
velocity of the vortex always has a direction opposite to 
the transport current which in turn results that the Hall effect
in the vortex state will have a sign which is opposite to that
of the normal state.
We should note here that equation (\ref{guidcen3}) is not original 
at all. The major contribution of this work is its field theoretical
derivation and its interpretation. Indeed, by multiplying both sides
of (\ref{guidcen3}) by $\, 2 \pi N \, $ and taking the cross product 
with $ \, \hat e_z \, $ we can write the latter in the form 
\begin{equation}
2 \pi N \, {\bf V}_L \times \hat e_z  \,=\, -2 \pi N \, {\bf J}_0 \times 
\hat e_z  \,-\, \eta  {\bf V}_L
\label{Magnuseqs2}
\end{equation}
\noindent
All the terms above appear in the most common phenomenological theories
of vortex motion \cite{rPhenom}. The left hand side of (\ref{Magnuseqs2})
is the familiar though controversial Magnus force  while the
first term of the right  hand side is the so-called Lorentz force
\cite{rLorentz},
or according to some other authors \cite{rThou} an inseparable part
of the Magnus force. Finally, the last term of the right  hand side 
of (\ref{Magnuseqs2}) is like the viscous drag force of the
Bardeen-Stephen (BS) model \cite{rPhenom}. However, one should notice 
that the Lorentz force comes with opposite sign in eq. (\ref{Magnuseqs2})
with respect to what is commonly accepted in the literature,
a difference 
which is of critical importance.  In fact, the sign inversion of the 
longitudinal part of the vortex velocity in our analysis, has its 
origin in the sign change of the Lorentz force.
The reason for the latter change is in the way we incorporate the external
current in our model.  We think of the current as a totally substantive
object which interacts with the vortex. It is like a vortex moving in a 
plane  under the influence of a current which flows in a parallel plane 
just above the vortex plane, instead of a current flowing in the
plane of the vortex and formed by the same carriers as those of the
vortex \cite{rPhenom}. At first sight the second approach looks more natural.
However, if we think of the vortex as a 3-dim object i.e. as
a flux tube formed in a 3-dim superconducting film and interacting with 
supercurrent which is formed only on the top and the bottom
of the film, our approach becomes more plausible and realistic.

\section{Numerical Results}
   
Our next objective is the numerical investigation
of the dynamical behavior  of the vortices in the presence of an
external electric current and dissipation. 
Our computational techniques are described in detail elsewhere
\cite{rST2}, so here we present only a brief overview.

To simulate the motion of a vortex, we first determined numerically 
the static profiles of the condensate and gauge fields characterizing 
a vortex of winding number $N \,$ \cite{rST}. 
With these in hand, we laid down on the lattice a configuration 
of a vortex centered at $(x,y) = (2,0)$ at time $t=0$. In order to 
maintain (as much as possible) the symmetries of the continuous system
in its discretized form, we have resorted to techniques from lattice
gauge theory \cite{rST2}, \cite{rMCarlo}.  The degrees of freedom were 
discretized on a spatial lattice so as to maintain exact (lattice) 
gauge invariance. 
The time evolution was 
implemented by a finite difference leapfrog method using equations
(\ref{rescaledeqs}) with the second one modified  to
\begin{equation}
{1\over \beta}\dot E_i=\epsilon_{ij} \partial_j B - J^s_i - J_i^{ext}
+\, C_d  \epsilon_{ij} \partial_j \dot B
\label{finaleqs}
\end{equation}
\noindent
in order to incorporate the effect of an external current and dissipation.
The gauge freedom of the equations of motion was eliminated by 
imposing the temporal gauge $\, A_0\,=\,0$.
The external current $\; {\bf J}^{ext} \;$  was taken along the
x-direction and uniform throughout the whole plane i.e. 
${\bf J}^{ext}=J_0 \hat{\bf e}_x$.

In any numerical calculation where partial derivatives are involved,
the imposition of appropriate boundary conditions is a very delicate task.
Here, the presence of an incoming and an outgoing external current at
$\pm$ x-infinity, made this issue even more complex and forced us 
to use different boundary conditions at $\, x$ and $\, y$-boundaries.
At the $\, y-$boundaries of the film we imposed free  
boundary conditions by setting the covariant derivative in
the normal to the boundary direction equal to zero ($D_y \Psi
\,=\, 0$). There are more than one ways to impose such a condition,
and our choice was to set $\,\partial_y \Psi=0$ and  $\, A_y=0$. 
Also, in order to get vanishing magnetic field at the $\, y-$boundaries
we set $\,\partial_y A_x=0$.
At the $\, x-$boundaries we successfully imposed two different sets of 
b.c. There, in deriving the boundary conditions we took special
care in order to preserve the discrete gauge invariance of the system,
or equivalently the discrete version of the continuity equation, 
at the $x-$boundaries. Specifically, in analogy to the 
$\, y-$boundaries we set 
$\,\partial_x A_y=0$, $\, A_x=0$  and then, by imposing the 
continuity constraint we got $ \, {1 \over i} [\Psi^{\star} 
\partial_x \Psi]=-J_0$.
As a consequence of the above b.c. the value of the x-component
of the supercurrent at $\pm x$-infinity was fixed and equal 
to $\, -J_0$.  Even though, this is what we expect to happen away from the 
vortex, still the b.c. for $\, \Psi$ sounds quite
artificial and too constrained. Thus, we tried and finally used another 
more ``natural'' set of b.c.  Namely we set
$\,\partial_x A_x=\partial_x A_y=0$ and the constraint of 
preserving the continuity equation yielded $\,D_x^2 \Psi=0$ as a b.c.
for $\Psi$. 
Using both sets we got essentially identical results, and  these
have been exhaustively checked to be free of any boundary contributions.

The simulations were done on a square lattice of $\, 201 \times
201 \,$ sites with a lattice spacing $\, \alpha \,=\, 0.15$.
The width (diameter) of the vortex was typically of the order of
$\, 2\, $ in rescaled length units. The finite time step was chosen to be 
much smaller than the lattice spacing typically of the order 
$\, 10^{-3} \, $.  To test our results
we ran simulations in bigger lattices $ 401 \times 401 $ with the same
or smaller lattice  spacing, say $\, \alpha \,=\,0.1$, and the results 
obtained were all perfectly consistent.
All our simulations were
performed on various HP workstations at the University of Crete. 
A typical run of duration
$T \approx 100$ time units, with  $\Delta t\,=\, 0.002 $
on a $201 \times 201$ lattice needed about 15 hours of CPU time
on a HP-735 machine.

Apart from the existence of the external current and the new boundary 
conditions, the algorithm we used here was identical to the one previously
used in the numerical study of a vortex pair dynamics \cite{rST2}.
There, it turned out that the algorithm was extremely accurate. Here,
as a sort of calibration of the algorithm, and mainly to avoid any 
systematic contributions from the b.c. we initially tested it in a
system where the result is known analytically, namely at the 
trivial sector $\,N=0$. There, it is easy to see that the field 
configuration $\Psi({\bf x},t) = exp[-{i \over 2} J_0^2 t],
A_x({\bf x},t) = J_0, A_y({\bf x},t)=0=A_0({\bf x},t)$
is a solution of the equations of motion in the presence of the 
external current with ${\bf E}({\bf x},t) = 0 = B({\bf x},t)$
and ${\bf J}^s({\bf x},t) = -{\bf J}^{ext}$.
 We ran a preliminary simulation using as initial configuration the 
trivial vacuum $\, \Psi=1, A_i=0=A_0 \,$ and turning on the external 
current at $\, t=0$.  Our naive expectation was to see the system 
relaxing to the vacuum-current solution described above or to some
gauge transform of it. Yet, after a short transient period the system 
dynamically relaxed to a time-dependent configuration where both the 
electric field and the supercurrent oscillated vividly 
around their mean values $ <{\bf E}({\bf x})> = 0$, $<{\bf J}^s({\bf x})> 
= -{\bf J}^{ext}$. These oscillating modes could be attributed to the 
abrupt turning on of the external current. To eliminate them, we 
introduced dissipation in the system and in a subsequent run we saw the
system relaxing to the vacuum-current solution within a $10^{-4}$ 
accuracy. (Note that we introduced here a dissipation term  of the
form $\, \dot E_i = -C_d' E_i \,+.... $ which does not meet all the 
criteria mentioned in the previous section and thus it is not 
appropriate in the vortex sector). The remarkably accurate convergence
of the initial configuration to the vacuum-current solution provides 
a strong confirmation for the accuracy and the reliability of our algorithm.
A byproduct of these runs is the conclusion
that the response of the ground state 
to the application of an external electric current is the formation
of a supercurrent which in average is equal and opposite 
to the latter. In fact, it is reasonable to assume that even in the 
$N \ne 0$ sector, away from the vortex, the system will response 
in the same way.

We switch now to the study of the vortex sector and we particularly
consider the $N = 1$ sector.
To study the dynamics of the vortices we carried out numerous simulations
and the results  confirmed with quite impressive accuracy 
the predictions of the theoretical analysis. We also experimented with
the values of the parameters $\kappa$,
$\beta$ and the dynamics of the vortices showed little sensitivity
to those values. Indicatively we quote here the results of some
simulations for the specific choice $\kappa = 2 $ and $\beta = 1$,
which belong to a parameter regime that we believe to be appropriate 
for the description of the physics of type II superconductors \cite{rST2}.
Similar results though, were obtained for a large variety of the values
of the parameters.

We performed a series of numerical experiments for various values
of the dissipation constant $C_d$ and for a fixed value of the
external current $\, J_0 \, = \, 0.025 $. 
The total duration of each simulation was $240$ time units.
The corresponding trajectories of the guiding center  of the vortex 
are displayed in fig.2.  As arises from the plot, in the absence 
of dissipation ($\, C_d = 0$), the guiding center  of the vortex 
performs a rectilinear trajectory along the  negative x-direction. 
Moreover, it is displaced by $\, 6 \,$ space units (i.e. three times 
its diameter) in $240$ time units, which corresponds to a mean velocity  
$\, {\bf V}_L = - J_0 \hat e_x$. This behavior verifies the theoretical
prediction (\ref{guidcen2}) qualitatively  as well as quantitatively.

   To arrive to eq. (\ref{guidcen2}) we assumed that
the external current is homogeneous throughout the whole space.
This assumption may take a weaker form; instead of a current occupying 
the whole space, we may introduce one which has a strip shape, i.e.
it takes a non-zero constant value inside a strip of certain width,
while it vanishes in the outside region. In principle, when the width 
of the strip is much larger than the size of the vortex, the vortex 
essentially realizes a homogeneous current  all over space 
and responds accordingly. 
This assumption was tested and verified in our study.
Specifically, we repeated our runs for $\, C_d=0 \,$ using
an external current of the form
\begin{equation}
      J^{ext}_y \,=\,0 \hskip 2.5cm
      J^{ext}_x \,=\, J_0 \,f(y)   \label{stripcurrent}
\end{equation}
\noindent
     where the function $f$ is approximately equal to unity over a
     strip of width $2L$ and drops to zero very quickly outside 
     that strip. A function which meets the above description is
\begin{equation}
f(y) \,=\, e^{-(|y|/L)^n}    \label{cutoff}
\end{equation}
\noindent
especially for large $n$.
The numerical calculations showed that any strip current 
with $\, L \geq 6 \,$ has the same effect on the motion of the vortex 
to that of a homogeneous current occupying the whole space. Quite 
surprisingly, even  
for narrower strips, $\, L \leq 6 \,$ the trajectory of the vortex 
remains identical while its speed reduces. Indicatively, we found
the ratio $ - {V_x \over J_0} \,=\,0.89$, $0.95$ and $0.98 $ 
for $L \,=\,3 $, $4$ 
and $5$ respectively.

For non-zero values of $C_d$, after a small transient period
the vortex relaxes to an almost rectilinear motion at an 
angle to the external current different from $\, 180^{\circ} $.
Indeed, the deflection angle $\delta$ between the trajectory 
of the vortex and the direction of the current takes values
from $\, 180^{\circ} \, $ to $\,125^{\circ} \, $ with $\delta$ decreasing
as $C_d$ increases. As arises from the lengths of the trajectories
in fig.2, the measure of the velocity of the vortex
is also a decreasing function of $C_d$ in accordance to the physical
interpretation of dissipation.

An important issue to check is to which extent the motion of the vortex 
meets the assumptions we adopted in section III while deriving 
eq. (\ref{goldenrule}). 
To see whether a steady state is eventually reached, we plot in fig.3
the time evolution of the $\, x\,$ and $\, y$-components of the drift 
velocity of the vortex for $\, C_d \,=\,4$.
There, we see that initially ($\,t=0 \,$),  $\, V_x=-0.025=-J_0 \, $ 
and  $\, V_y=0 \, $ as if there was no dissipation at all.
As the vortex moves on, dissipation turns on and its effect results
in a non-zero transverse component of the vortex velocity.
After a small transient period ($\, t = 0 - 4 \,$), the velocity of 
the vortex sets in an oscillating mode around a constant mean value. 
Strictly speaking the vortex does not seem to develop a steady state,
but it is reasonable to assume that the contribution of the
oscillating part is auto-canceled in average, and thus in a wider
sense we can assert that the vortex finally relaxes to a steady state.
Also, a detailed examination of successive level contours of the 
energy density establishes that the vortex moves quite coherently
and retains its initial shape during the evolution of the simulations.
It is thus quite interesting to question whether the trajectories 
shown in fig.2 satisfy eq. (\ref{goldenrule}).
To calculate the deflection angle and the (mean) velocity
$\, {\bf V_L} \, $ of the vortex we process the numerical data 
so as to linearize the slightly wavy trajectories of fig.2 and 
we cut out the initial part of the data which corresponds 
to the ``transient period''.
After the relevant calculations we tabulate the results in table I, 
where it is demonstrated that relation (\ref{goldenrule}) is satisfied
with quite impressive accuracy.

\vspace*{.3cm}
\centerline{ \bf Table I}
\vspace*{-.1cm}

\begin{center}
\begin{tabular}{||c|c|c|c||} \hline
               \hspace{0.2cm} $C_d$                 \hspace{0.2cm} &
               \hspace{0.2cm} $ -cos \delta$    \hspace{0.2cm} &
      \hspace{0.2cm} ${V_L \over   J_0}$       \hspace{0.2cm} &
   \hspace{0.2cm} $-{cos \delta \over V_L/J_0}$ \hspace{0.2cm}
               \\ \hline
               $0.$    &  1.000   & 1.000    & 1.000   \\ \hline
               $1.$    &  0.984   & 0.981    & 1.003   \\ \hline
               $2.$    &  0.942   & 0.934    & 1.009   \\ \hline
               $4.$    &  0.816   & 0.803    & 1.016   \\ \hline
               $8.$    &  0.579   & 0.582    & 0.995   \\ \hline
\end{tabular}
\end{center}

\vspace*{.2cm}
It is of crucial importance for the present study to determine 
to which extent the details of the motion of the vortex follow the 
motion of its guiding center. To investigate the latter we must 
somehow define the position of an extended object such as the vortex.
Thinking of the  vortex as an energy lump and given that this energy 
lump moves coherently retaining its initial axially symmetric shape,
we believe that the maximum of the energy density (MED) 
of the vortex configuration accurately describes the real
position of the vortex. Therefore, in our calculations we also keep 
a record of the location of the MED of the vortex configuration as 
time evolves.  Note that there are at least two alternative definitions
of the position or the center of the vortex, namely the maximum of the 
topological density  or the position where $\, \Psi \, $ vanishes.
All three of them yield similar results so here we present only 
data for the MED. 
The trajectories of the MED of the vortex (solid line)
and of the guiding center (dashed line) of the vortex for 
$\, C_d=0 \, $ and for various values of the intensity $\, J_0 \, $ of 
the external current are plotted in fig.4.
We see that while the guiding center simply drifts in the negative 
x-direction, the  motion of the MED is more complicated.
In average it follows
the motion of the guiding center, but  its trajectory is modulated
by an oscillatory pattern.  This modulation is not a numerical effect 
but is an inherent characteristic of vortex dynamics.
We  have already encounter similar oscillating patterns in an earlier
work while studying the dynamical evolution of a pair of vortices \cite{rST2}.
These patterns are reminiscent of the motion that electrically charged 
particles perform in the presence of a magnetic field, which happens to
be the prototype physical system which exhibits
Hall behavior.  Borrowing the 
terminology from the latter, we will  refer to this finer motion with 
the name ``cyclotron''.  As in the original case, the amplitude of the 
cyclotron motion varies with the parameters of the problem. 
Along with the amplitude, the importance of cyclotron motion also varies.
In the extreme limit where the amplitude of the cyclotron motion is very
large in comparison with the length scale of the problem we study, 
the whole picture of the dynamical behavior  of vortices, as this is 
determined from the equation of motion for the guiding center alters 
dramatically. Thus, it is important to determine the way the parameters 
of the model affect  cyclotron motion. Our numerical investigation
revealed a systematic relation between the parameter $\beta$, the 
value of the external current $J_0$, and the amplitude of the
oscillating patterns.  In short, the cyclotron motion is amplified 
when $\beta$ 
decreases as well as when $J_0$ increases. The dependence on $J_0$ and 
$\beta$ is demonstrated in fig.4 and fig.5 respectively. Note that
in the runs presented in fig.4 $\, \kappa=2.$ and $\, \beta =1. \,$
while in those in fig.5 $\, \kappa=1.5$ and $\, J_0=0.03 $. 
Here we stress once more the analogy between the motion of the 
vortex in this model and the Hall motion of charged particles. The increase 
of $J_0$ here is equivalent to the increase of the external electric field
in the charged particle system and the effect on the cyclotron motion
is the same in the two cases. Also the increase of $\beta$ is 
physically equivalent to the increase of the dielectric constant.
The analogue in the particle system is to decrease the coupling of the
particle with the electric field, without affecting its coupling with
the magnetic field. Again the consequence is similar in the two systems.

Finally we consider the case where an initially applied
external field is abruptly turned off.  The response of the vortex
to such a ``blackout'' is of obvious interest. 
Hall equation (\ref{Halleqs})
implies that in the absence of external
forces the guiding center of the vortex
is conserved i.e. the vortex is pinned.
In other words, while the vortex  moves at a constant speed
just before the current is turned off, it abruptly freezes at the 
position where it is found at the time we switch off the current.
Still there is one question to be answered, namely, how the location
of the MED -which in principle does not coincide with the guiding center-
will evolve.
After all the guiding center is an abstract notion which represents
the ``mean position'' of the vortex, while its real position in space
is associated with the distribution of the energy density.
One could possibly assume that after the pause of the current
the vortex will  reorganize itself and will finally relax to a
configuration where the MED coincides with the pinned guiding center.
However, the numerical results lead to a completely different
picture. 

In fig.6 (a), (b) we display the results of a simulation
of a total duration of $125$ time units where an external current
is on only up to time $t\,=\,t_{crit}\,=\,40$.
The trajectories of both the guiding center (solid line)
 and the MED (dashed line) are plotted. Plot (a) shows the
trajectories from $t\,=\,0$ up to $t\,=\,t_{crit}$ while plot (b)
displays the trajectories till $t\,=\, 125$.
As it is shown there, after the current is turned off,
the guiding center indeed remains fixed at the point
where it was found then, while the maximum of the energy density
sets in a circular motion around the location of the guiding center
with a radius equal to its distance from the guiding center at
the time the current was turned off.
 The picture described here, has a striking analogy with the Hall effect.
 A guiding center can be also introduced in the case of the
two-dimensional electron motion in a uniform magnetic field $B$ \cite{rPTa}
which again can be interpreted as  the ``mean position'' of the electron.
When an external electric field is applied the electron
sets in a cycloid motion along the Hall direction while its
guiding center follows a rectilinear orbit along the same direction.
What is more, when the electric field is turned off, the electron,
sets in a circular motion  around its guiding center which rests.

\section{Discussion}

In this paper, a time-dependent Ginzburg-Landau model for a complex
scalar field coupled to electromagnetism has been studied mainly
numerically.
Dissipation was successfully incorporated in the model and its effect on 
the motion of the vortex was analytically determined.
The results of the numerical study were in accord to 
the Hall analogy advocated in ref. \cite{rPTb} based on the derived
unambiguous conservation laws. Furthermore, the results confirmed
with impressive accuracy earlier theoretical predictions concerning 
the speed and the direction of the velocity of the vortex.
In short, it was shown that under the influence of an external electric 
current a vortex drifts in a direction opposite to the current while 
in the presence of dissipation it deflects in a direction ranging from 
$90^{\circ}$ to $180^{\circ}$ with respect to the current.
 
An important feature which is worth emphasizing is the cyclotron motion
i.e. the oscillating patterns in the trajectory of the vortex remarkably 
similar to those encountered in the cycloid motion of an electron in 
the standard Hall effect. Such patterns have been already observed in the 
motion of vortex pairs \cite{rST2} and in the motion of magnetic bubbles 
in ferromagnetic media \cite{rPZ1}. 
It seems, that cyclotron motion is a generic feature of solitons
which exhibit Hall behavior.  
According to reference \cite{rST} there is a whole class of 
field theories
whose solitons are expected to exhibit Hall behavior and  
among them there are some interesting variations of the present 
model \cite{rManton}, \cite{rChern}.
In some sample runs in these 
models we did encounter cyclotron motion which we consider as a 
strong indication that indeed these systems exhibit Hall behavior.
Note that in the framework of  collective coordinate schemes
like those invoked in \cite{rManton}, \cite{rPZ2} it is
not possible to detect cycloid patterns, because in the adiabatic
limit the amplitude of cyclotron motion becomes negligible, a fact 
which is supported by the results displayed in fig.4.

At a phenomenological level, we have presented arguments
(mainly by reformulating previous results), which are quite 
encouraging for the relevance of the model to the physics of 
the superconductor. In particular, we have shown that Hall 
equation (\ref{Halleqs}) leads naturally to the introduction 
of Magnus force (\ref{Magnuseqs}) in the equation of motion 
for the vortex. Furthermore, we argued that the Magnus force has 
electromagnetic origin due to the interaction of the moving magnetic 
flux with the positive ions of the background lattice. 
Finally, we have derived a variation of the Nozieres-Vinen
equation for the motion of a vortex (\ref{Magnuseqs2})using plain 
field theoretical analysis.

We have also exhibited analytical considerations and
numerical results, which suggest that in the presence of an
external electric current the vortex drifts against the current
implying a possible link with the opposite sign Hall effect
\cite{rOpposite}.
Yet, as we mentioned in section IV, the way we incorporate
the electric current in the model plays crucial role to the
derivation of that result. Clearly the next step is to test the
predictions of the model at hand against more realistic experimental 
situations.  One should find a more natural way to 
introduce the electric current in the specimen. Also one should abandon
the 2-dim reduction and study the issues presented here in thin films
with finite thickness. In such a study \cite{r3dim}, preliminary 
results imply that  Hall 
behavior is also exhibited in the motion of magnetic flux tubes
which are probed by a surface external current 
in a 3-dim film i.e. a current which is non-zero only at the upper 
and the lower layers of a 3-dim grid. 
At the experimental front, one might try to mobilize vortices not
by applying an electric potential on the specimen, but by introducing 
an electric current in a parallel plane just above the specimen.

{\bf Acknowledgments}

I would like to thank Professors T.N. Tomaras and N. Papanicolaou for
several helpful discussions. I would like to acknowledge the 
hospitality of the Edinburgh Parallel Computing Center where 
part of this work was performed.  This research was supported
in part by the EU grant CHRX-CT94-0621  and by
the Greek General Secretariat of Research and Technology grant 
No $95 \Pi {\rm ENE}\Delta 1759$.

\vfill

\newpage

\vskip 2cm
\hskip 0.5cm {FIGURE CAPTIONS}

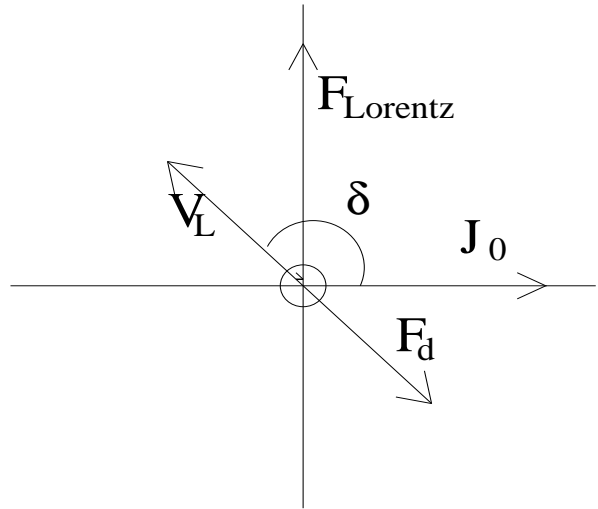
{\bf Figure 1.}  A plot of the forces acting upon a vortex which
moves with constant velocity $\, {\bf V}_L \,$ in the presence of a
uniform external current  $\, {\bf J}_0 $ and dissipation.

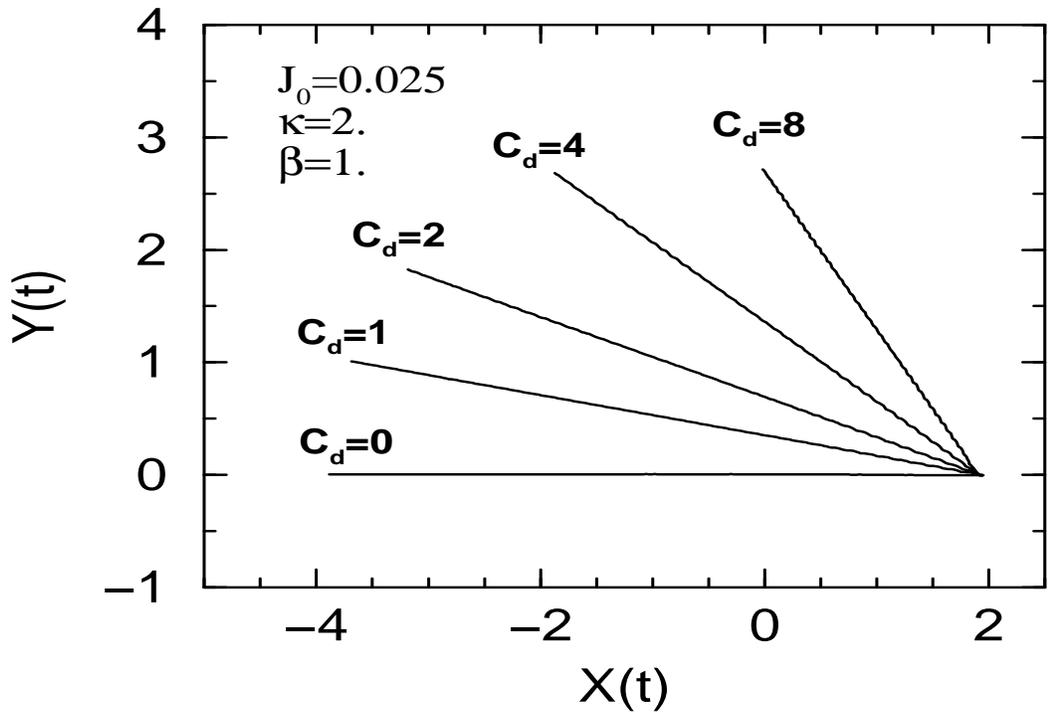
{\bf Figure 2.}  The trajectory of the guiding center of the vortex
under the influence of an external electric current and dissipation.
The several lines correspond to different values of the friction
parameter $\, C_d$.

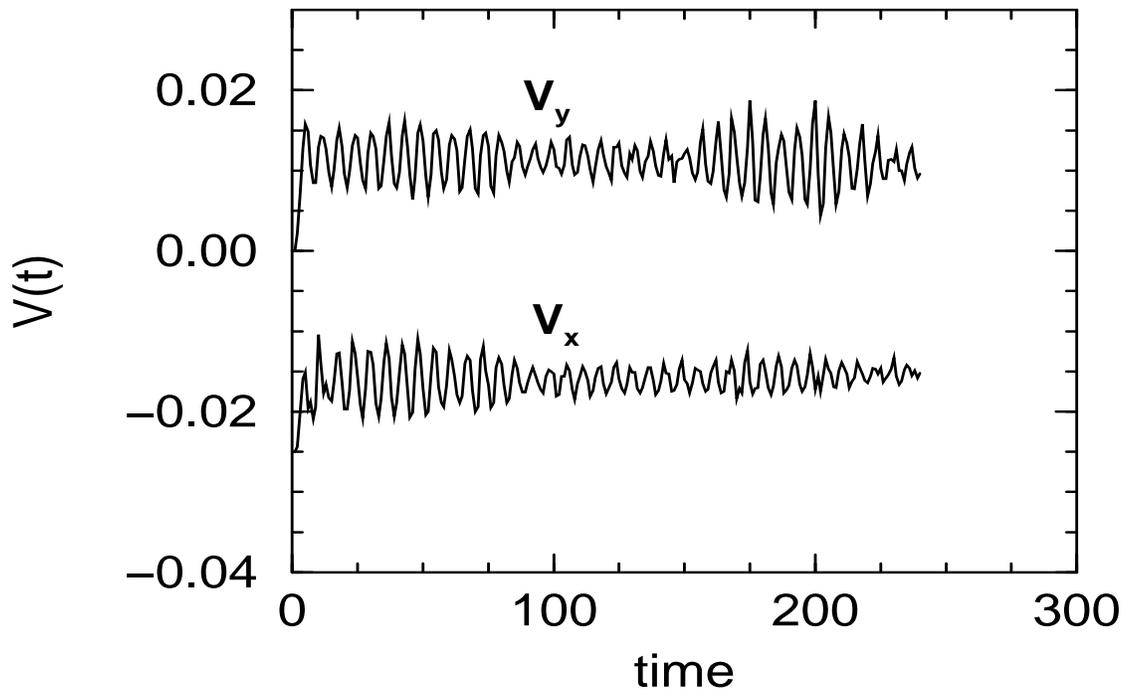
{\bf Figure 3.}  The drift velocity ${\bf V} = (V_1,V_2)$ of the
guiding center of the vortex for $\, C_d\,=\,4$.

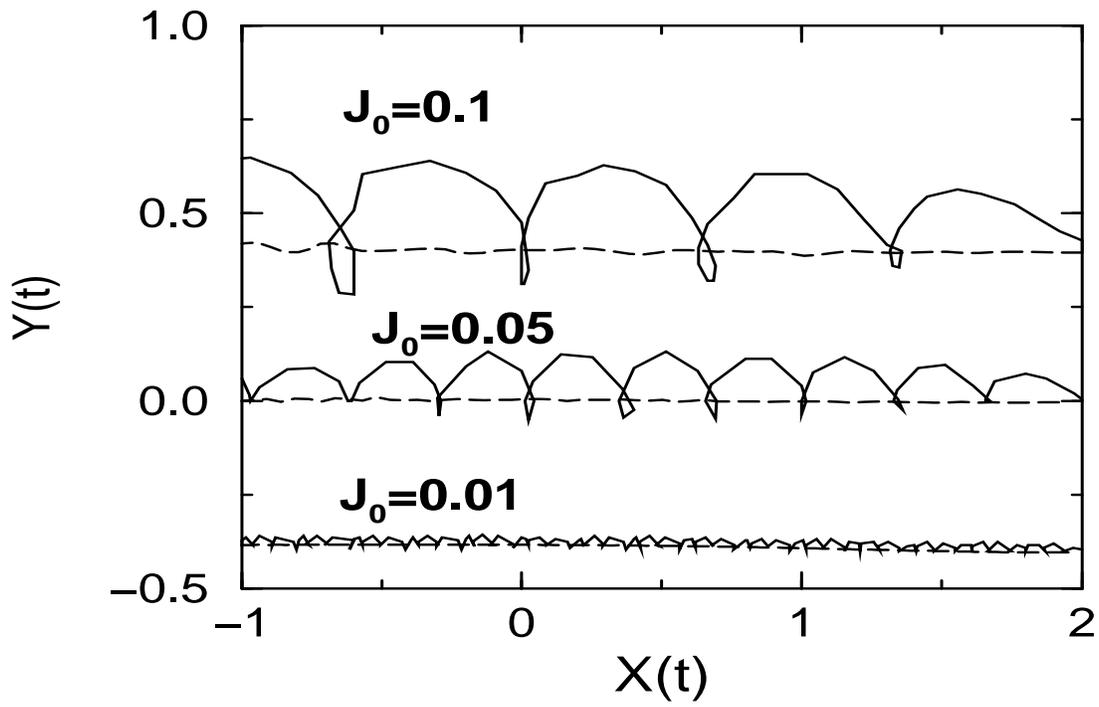
{\bf Figure 4.}  The motion of the vortex as determined by
its guiding center (dashed line) and the location of the maximum of
the energy (solid line) for various values of the external current.

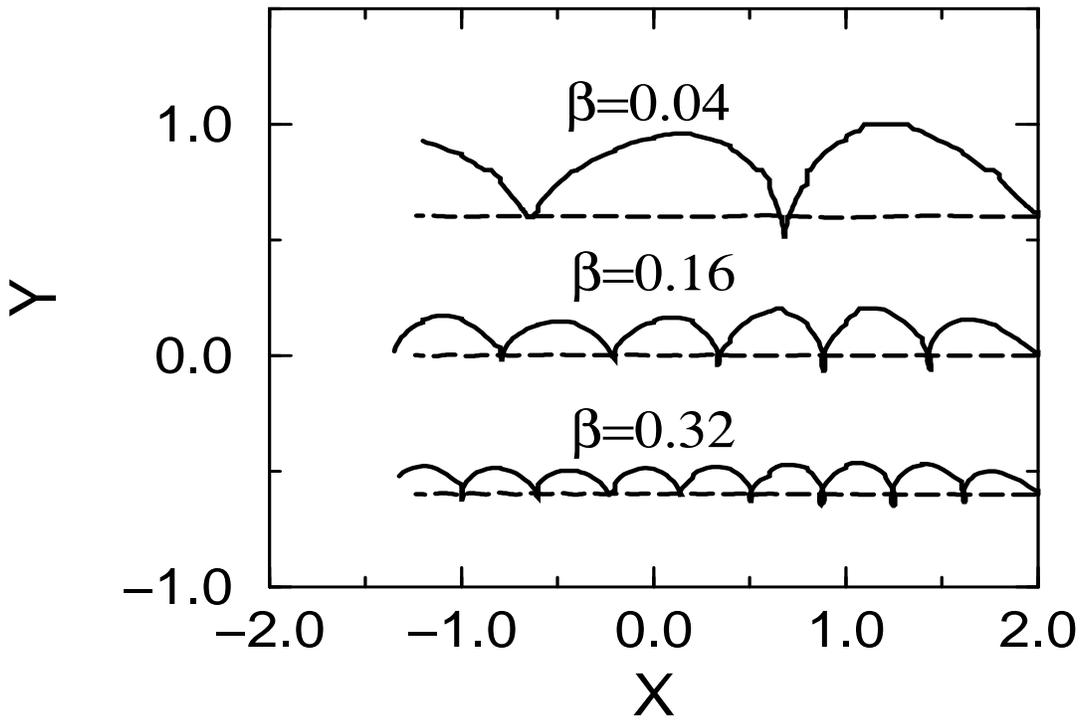
{\bf Figure 5.}  The motion of the vortex as determined by
its guiding center (dashed line) and the location of the maximum of
the energy (solid line) for various values of the parameter $\beta$.

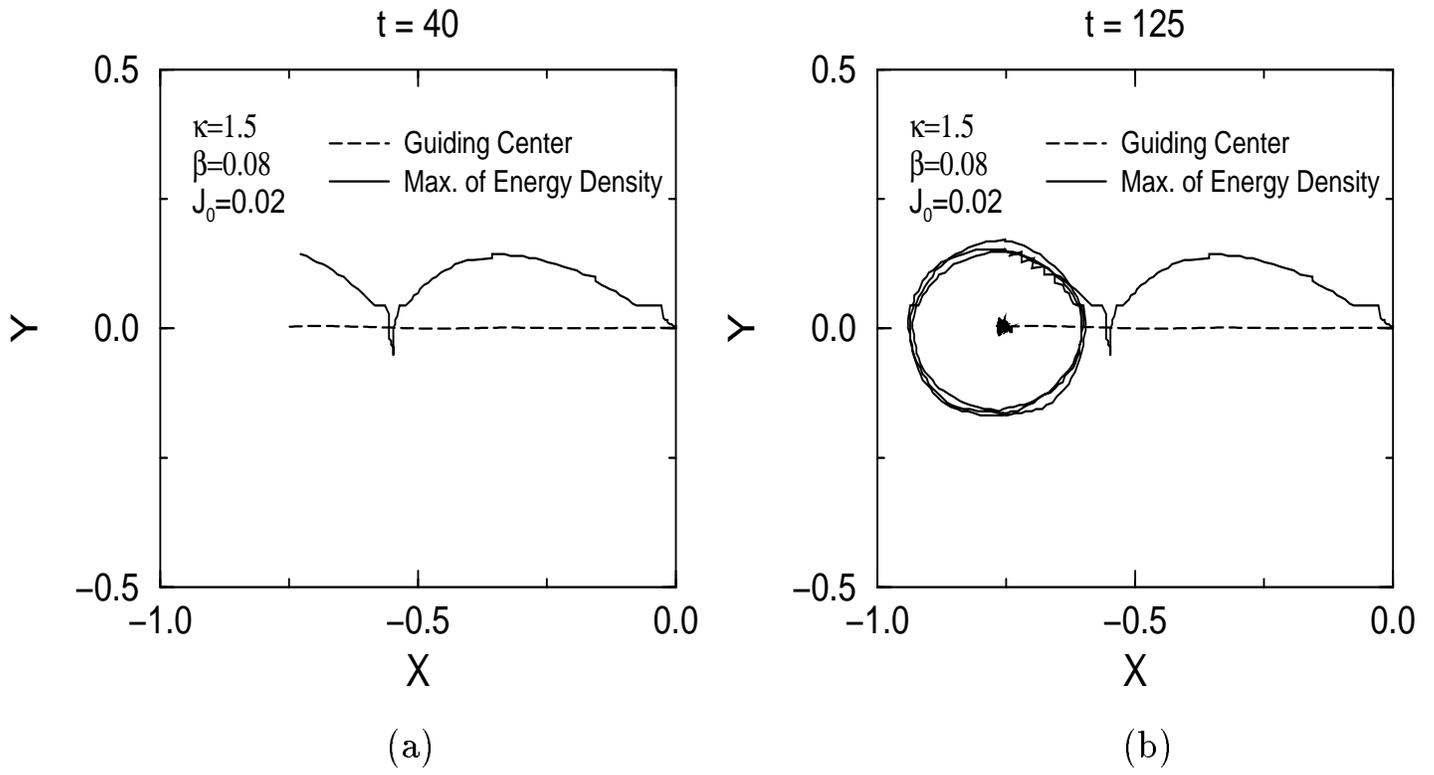
{\bf Figure 6.}  The pinning of the vortex.
(a) the trajectory of the vortex while the current is on, and
(b) the subsequent evolution after the current is turned off.

\end{document}